\documentclass[prl,twocolumn,showpacs,amsmath]{revtex4}
\usepackage{graphicx}
\usepackage{dcolumn}
\usepackage{bm}

\begin{document}

\title{Supernarrow spectral peaks near a kinetic phase transition in a driven, nonlinear micromechanical oscillator}

\author{C. Stambaugh}
\author{H. B. Chan}
\email{hochan@phys.ufl.edu}
\affiliation{Department of Physics, University of Florida, Gainesville, FL 32611}

%
\begin{abstract}
We measure the spectral densities of fluctuations of an underdamped nonlinear micromechanical oscillator. By applying a sufficiently large periodic excitation, two stable dynamical states are obtained within a particular range of driving frequency. White noise is injected into the excitation, allowing the system to overcome the activation barrier and switch between the two states. While the oscillator predominately resides in one of the two states for most excitation frequencies, a narrow range of frequencies exist where the occupations of the two states are approximately equal. At these frequencies, the oscillator undergoes a kinetic phase transition that resembles the phase transition of thermal equilibrium systems. We observe a supernarrow peak in the power spectral densities of fluctuations of the oscillator. This peak is centered at the excitation frequency and arises as a result of noise-induced transitions between the two dynamical states.  
\end{abstract}

\maketitle


Periodically driven nonlinear systems often display multiple coexisting dynamical states under sufficiently strong driving fields. The presence of fluctuations enables these systems to occasionally overcome the activation barrier in phase space, resulting in transitions between the dynamical states \cite{1}. Noise-induced switching has been studied experimentally in a number of dynamical nonlinear systems that are far from equilibrium, including parametrically driven electrons in a Penning trap \cite{2}, radio frequency driven Josephson junctions \cite{3} and micro- and nano-mechanical oscillators \cite{4,5,6}. When the noise is weak, the escape rate  $\Gamma_i$ out of state $i$ ($i = 1$ or $2$) depends exponentially on the ratio of an activation barrier $R_i$  and the noise intensity $I_N$:
	\begin{equation}\label{eq:1}
        \Gamma_i  \propto e^{-R_i/I_N}.
\end{equation}		
 $R_i$ typically depends on system parameters such as the driving frequency and amplitude, as well as the shape of the power spectrum of the noise. The ratio of the populations of the two dynamical states is given by: 	
	 \begin{equation}\label{eq:2}
        w_1/w_2  \propto e^{(R_2-R_1)/I_N}.
\end{equation}	
As a result of the exponential dependence of the population ratio on the difference in the activation barriers, the system will be found in either state $1$ or state $2$ with overwhelmingly large probability over most of the parameter space \cite{1,7}. The occupations of the two states are comparable only over a very narrow range of parameters. This behavior bears close resemblance to systems in thermal equilibrium with two phases such as liquid and vapor. Such thermodynamic systems are usually in either one of the two phases and only at the phase transition will the two phases coexist. Even though driven, nonlinear systems are in general far from equilibrium, theoretical works predicted that a similar kinetic phase transition would occur under the appropriate conditions \cite{1}. Like thermodynamic systems, fluctuations increase significantly when these non-equilibrium systems undergo phase transitions. A range of phenomena, including the appearance of a supernarrow peak in both the susceptibility and the spectral density of fluctuations \cite{1,8,9}, is expected to take place.

\begin{figure}[ht]
\includegraphics[angle=0]{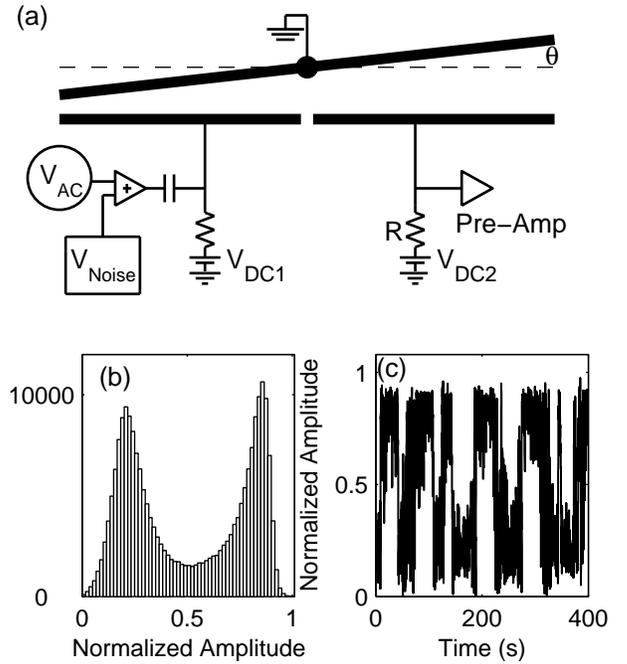} 
\caption{\label{fig:1}(a) A cross-sectional schematic of the micromechanical torsional oscillator with electrical connections and measurement circuitry (not to scale).  (b) Histogram of the oscillation amplitude of the oscillator near a kinetic phase transition showing equal occupation of the two states. (c) Time trace of the normalized oscillation amplitude near a kinetic phase transition, demonstrating transitions between the two states.   
}
\end{figure} 

In this letter, we measure the spectral densities of fluctuations of an underdamped nonlinear micromechanical torsional oscillator near the kinetic phase transition where the populations of the two attractors are comparable. The most prominent feature in the fluctuation spectrum is a narrow peak centered at the frequency of the periodic excitation. We demonstrate that this narrow peak is associated with noise-induced transitions between the two attractors. The width of the peak varies linearly with the transition rate and is more than a factor of 10 smaller than the natural line width of the resonance in our experiment. Away from the kinetic phase transition, the intensity of the peak decreases exponentially. Apart from the narrow peak, we also observe smaller, much broader peaks in the spectrum that are associated with fluctuations within each attractor. These broad peaks are present for all driving frequencies within the hysteresis loop and their dependence on the noise intensity is distinctly different from the narrow peaks at the kinetic phase transition.

A micromechanical oscillator, consisting of a moveable polysilicon plate supported by two torsional rods, was used to take the measurements in our experiment. Two fixed electrodes are located beneath the movable polysilicon plate, one on each side of the torsional rods. A voltage $V$ applied to one of the electrodes generates an electrostatic torque that excites the torsional oscillations. The other electrode is used to capacitively detect the oscillations. All measurements were taken at a temperature of $70\;K$ and pressure $< 10^{-7} \;torr$.  The driving voltage $V$ is a sum of a biasing dc voltage, a periodic ac voltage with frequency  $f_d = \frac{\omega_d}{2\pi}$ and random noise. The oscillator, driven simultaneously by a periodic excitation and random noise, is well-described by the equation 
 	\begin{equation}\label{eq:3}
         \ddot{\theta} + 2 \gamma  \dot{\theta}  + \omega_{0}^{2}\theta  + \beta \theta^3  = E\sin(\omega_d t) + n(t), 
\end{equation}
where  $\omega_0$ is the natural frequency for small oscillations, $E$  is the effective amplitude of the periodic excitation, $\gamma$  is the damping constant, $n(t)$ is the effective noise in the excitation and  $\beta$ is the coefficient of the cubic nonlinearity (an omitted term that is quadratic in  $\theta$ leads to renormalization of the nonlinear coefficients). Figure \ref{fig:1}a shows a schematic of this setup. A more thorough description of the setup and derivation of the equation of motion can be found in Refs. \cite{5,10}.

\begin{figure}[ht]
\includegraphics[angle=0]{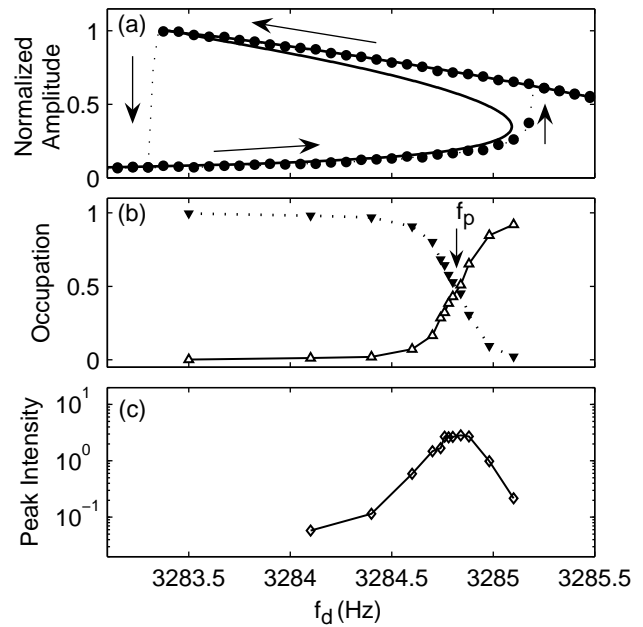} 
\caption{\label{fig:2}(a) Normalized frequency response of the oscillator (circles) fitted to a damped oscillator with cubic nonlinearity (solid line).  (b) Occupation of the two states versus frequency. At the kinetic phase transition ($f_p = 3284.8 \;Hz$), the occupation of the high amplitude state (upright triangles) and the low amplitude state (inverted triangles) are comparable. (c) Dependence of the intensity of the supernarrow spectral peak on the driving frequency $f_d$.  The intensity falls off exponentially as the driving frequency is moved away from $f_p$.}
\end{figure}  

	In the absence of noise and at small driving torque, the response of the oscillator corresponds to that of a damped harmonic oscillator with a resonant frequency of $3286 \;Hz$ and $Q \approx 8000$. When the driving torque is increased beyond a critical value, the frequency response becomes hysteretic due to the cubic nonlinearity. Within a range of driving frequencies, two stable dynamical states coexist. In the absence of fluctuations, no transitions occur between the two states. When sufficient noise is injected into the driving voltage, the oscillator switches between the two attractors. Over most of the hysteresis loop, the activation barriers $R_{1,2}$  for escape from the two states are significantly different. The hollow and solid triangles in Fig. \ref{fig:2}b shows the occupation of the high-amplitude and low-amplitude states respectively. On the low frequency side of the hysteresis loop, the occupation of the low amplitude state is considerably higher than the high amplitude state and the probability of finding the oscillator in the low amplitude state is close to unity. As the driving frequency decreases, the activation barrier  $R_1$ for switching out of the high amplitude states decreases. At the bifurcation frequency ($f_1  = 3283.3 \;Hz$), $R_1$  goes to zero and the low-amplitude state becomes the only attractor. In an earlier experiment \cite{5}, we showed that  $R_1$ depends on the detuning frequency ($f_d-f_1$) in the vicinity of the bifurcation point with a critical exponent of 3/2 in agreement with theoretical predictions \cite{1}. On the high frequency side of the hysteresis loop, a similar argument applies except that the high-amplitude state is the stable attractor.
	
	While the oscillator is predominantly in one of the attractors over most of the hysteresis region, there exists a small range of frequencies where the occupation of the two attractors is of the same order of magnitude. Figure \ref{fig:1}c shows the oscillation amplitude as a function of time at a driving frequency of $3284.8 \;Hz$ and clearly illustrates the system switching between the two states. The relative occupation of the two states at this driving frequency is deduced by calculating the area under the two peaks in the histogram of the oscillation amplitude (Fig. \ref{fig:1}b).
	
	Even though the nonlinear oscillator is a driven system that is far from equilibrium, the above behavior bears resemblance to thermodynamical systems at phase transitions when two phases coexist. A kinetic phase transition was predicted to occur in non-equilibrium systems when the populations of the two dynamical states are equal \cite{1}. An important feature associated with phase transitions is the large fluctuations arising from transitions between the two states. We perform a systematic study of the fluctuations of the nonlinear oscillator at the kinetic phase transition by examining the power spectral densities of the oscillator response. 
	
	To resolve features in the spectral densities of fluctuations, it is necessary to have high resolution in frequency. The maximum number of consecutive data points that can be recorded with our instrument limits the frequency resolution in calculating the fast Fourier transform (FFT) of the angular displacement  $\theta(t)$. Instead, we record the slowly varying envelope of the oscillations using a lock-in amplifier with a bandwidth of about $30 \;Hz$.  The response of the oscillator can be written as 
  				\begin{equation}\label{eq:4}
         \theta(t) = A(t)\cos(2\pi f_dt) + B(t)\sin(2\pi f_dt),
\end{equation}
where $A(t)$ and $B(t)$ are  the slowly varying amplitudes of oscillations in and out of phase with the driving torque at frequency  $f_d$. 
The spectral density of fluctuations is given by
 \begin{equation}\label{eq:5}
 \begin{split}
Q(f)& =\frac{1}{N}\sum_{\substack{\tau}}\sum_{\substack{t}}\Bigl[\bigl(A(t)-iB(t)\bigr) \Bigl.\\
& \Bigl. \ \times\bigl  (A(t+\tau)+iB(t+\tau)\bigr)e^{-i2\pi(f-f_d)\tau}\Bigr],
\end{split}
\end{equation}
where $N$ is a normalization constant.

\begin{figure}[ht]
\includegraphics[angle=0]{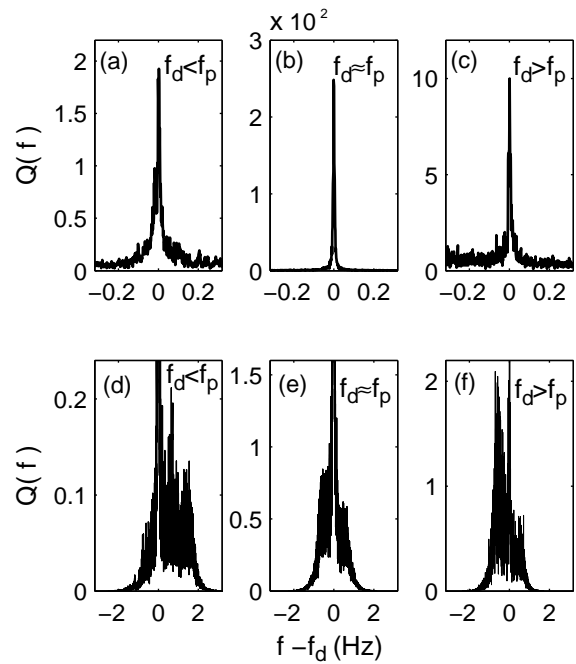} 
\caption{\label{fig:3}Power spectral density of fluctuations for three different drive frequencies $f_d$: (a) $3284 \;Hz$, (b) $3284.7 \;Hz$, (c) $3285.1 \;Hz$. Notice that the vertical scale of (b) is more than $20$ times larger than (a) and (c). In (d), (e) and (f) the axes have been rescaled to reveal the smaller and broader peaks in the fluctuation spectrums of (a), (b) and (c) respectively.    
}
\end{figure}

 In the absence of injected noise, oscillations occur only at the periodic driving frequency  $f_d$ and the measured spectrum consists of a delta function centered at $f_d$. When noise is added to the excitation, the spectral densities of fluctuations become dramatically different. Figure \ref{fig:3} shows the spectral density of fluctuations at $3$ different periodic driving frequencies. To focus on fluctuations about the ensemble average response, the delta function peak at $f_d$  obtained with no injected noise is removed from the spectrum. In other words, the data point at $f_d$  is omitted for each panel in Fig. \ref{fig:3}. Figure \ref{fig:3}b shows the spectral density of fluctuations at the driving frequency $f_d \approx f_p$  where the occupations of the two states are equal. The most prominent feature is a very sharp peak centered at the driving frequency. The width of this peak is a factor of $10$ smaller than the natural width of the resonance peak. This sharp peak is predicted to arise \cite{1} due to fluctuation-induced transition between the two dynamical states. Figures \ref{fig:3}a and \ref{fig:3}c shows the spectral density at two other driving frequencies that are comparatively far away from  $f_p$. As the periodic driving frequency is changed so that the oscillator moves away from the phase transition point, the sharp peak shifts accordingly to remain centered at the driving frequency. The area under the peak, however, drops significantly. Figure \ref{fig:2}c plots the area under the narrow peak as a function of periodic driving frequency, clearly demonstrating that the intensity of the supernarrow peak attains maximum at the kinetic phase transition and decreases exponentially as the occupation of one of the states exceeds the other and transitions between the states become less frequent.  
 
\begin{figure}[ht]
\includegraphics[angle=0]{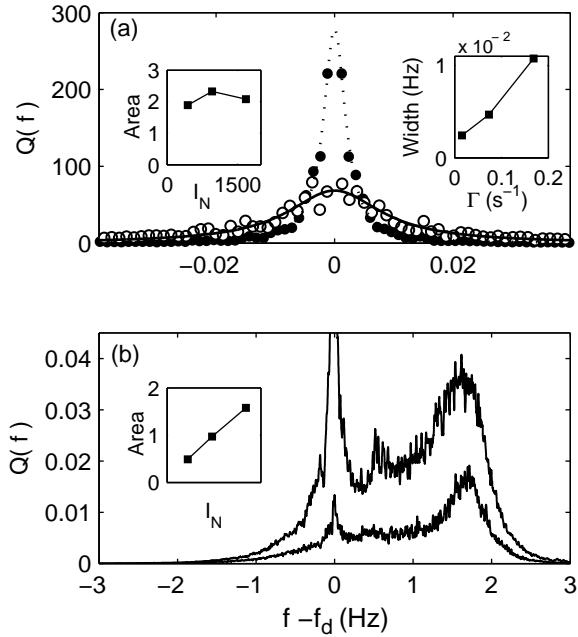} 
\caption{\label{fig:4}(a) The supernarrow peak for two intensities of injected noise at the kinetic phase transition. The injected noise intensity for the hollow circles is $4$ times greater than the solid circles. The dotted and solid lines are Lorentzian fits to the data. Left inset: Dependence of the area of supernarrow peak on injected noise intensity. Right inset: The peak width vs. transition rate. (b) The broad, small peak at two injected noise intensities that differ by a factor of 2, at $f_d = 3284 \;Hz$. Inset: Linear dependence of the peak area on injected noise intensity.  }
\end{figure}

Figure \ref{fig:4}a shows the behavior of the supernarrow peak with different noise intensities when the periodic driving frequency is held constant to maintain the oscillator at the kinetic phase transition. The noise intensity differs by a factor of $4$ between solid and hollow circles. Both sets of data are fitted well by Lorentzians \cite{8,9}. As the noise increases, the peak width increases and the peak height decreases. We found that the area under the peaks remains about constant, changing by less than $10 \%$ when the noise intensity changes by more than a factor of four (left inset of Fig. \ref{fig:4}a). The right inset of Fig. \ref{fig:4}b plots the peak width as a function of the sum of the transition rates $\Gamma_i$ out of each state, where the transition rates are determined directly from the residence time between transitions. The linear dependence of the peak width on the transition rate and the exponential decrease in the area of the peak away from the kinetic phase transition clearly identify the supernarrow peak with noise-induced transitions between the attractors. 

In addition to the supernarrow peak, there are other peaks in the fluctuation spectrum that are weaker and much broader than the supernarrow peak at the kinetic phase transition. These peaks, unlike the supernarrow peak, are present for all excitation frequencies within the hysteresis loop, as shown in Figs. \ref{fig:3}d and \ref{fig:3}f when the oscillator is far from the kinetic phase transition and in Fig. \ref{fig:3}e when the oscillator is at the kinetic phase transition. These smaller peaks represent characteristic frequencies of fluctuations about each dynamical state when the fluctuations are not strong enough to induce a transition over the activation barrier. In contrast to the supernarrow peak, the shape of these small peaks do not change considerably as the noise intensity increases (Fig. \ref{fig:4}b) and their area varies proportionally with noise intensity (inset of Fig. \ref{fig:4}b).

Systems in thermal equilibrium, such as a Brownian particle fluctuating in symmetric double-wells potentials, also exhibit narrow peaks in their fluctuation spectrum \cite{11}. The broadening of these peaks with increasing noise intensity leads to the well known phenomenon of stochastic resonance \cite{12}. We emphasize here the similarities and differences of our nonlinear oscillator to these systems in thermal equilibrium. First, our oscillator is far from equilibrium \cite{13,14,15} and is not characterized by free energy. It is bistable only if it is driven strongly enough and is monostable otherwise. Second, at the kinetic phase transition, the sharp spectral peak is centered at the driving frequency instead of zero frequency. The study of such critical kinetic phenomena in periodically driven nonlinear micro- and nano-mechanical oscillators could open new opportunities in tunable narrow band filtering and detection.

We thank M. I. Dykman and D. Ryvkine for useful discussions.


\end{document}